\documentclass[usenatbib]{mn2e}
\usepackage{epsfig}

\title[A Search for Planets Transiting the M~Dwarf Debris Disk Host, AU~Microscopii]
{A Search for Planets Transiting the M~Dwarf Debris Disk Host, AU~Microscopii}
\author[L.~Hebb et al.]{Leslie Hebb$^{1,2}$\thanks{E-mail: leslie.hebb@st-andrews.ac.uk}, 
Larry Petro$^{3}$, Holland C. Ford$^{2}$, David R. Ardila$^{2,5}$, Ignacio Toledo$^{6}$,
\newauthor
Dante Minniti$^{6}$, David A. Golimowski$^{2}$, Mark Clampin$^{4}$ \\
$^{1}$School of Physics and Astronomy, University of St Andrews, North Haugh, St Andrews KY16 9SS, UK \\
$^{2}$Department of Physics and Astronomy, The Johns Hopkins University, 3400 North Charles St, Baltimore MD 21218 \\
$^{3}$Space Telescope Science Institute, 3700 San Martin Drive, Baltimore MD 21218 \\
$^{4}$Spitzer Science Center, California Institute of Technology, Pasadena, CA 91125 \\
$^{5}$Department of Astronomy, Pontificia Universidad Catolica de Chile, Casilla 306, Santiago 22, Chile \\
$^{6}$NASA Goddard Space Flight Center, Code 681, Greenbelt, MD 20771. }

\begin{document}

%\date{Received 2007 February 19}
%
%\pagerange{\pageref{firstpage}--\pageref{lastpage}} \pubyear{2007}

\maketitle

\label{firstpage}

\begin{abstract}
We present high cadence, high precision multi-band photometry of the young, M1Ve, debris disk star, AU Microscopii.  
The data were obtained in three continuum filters spanning a wavelength range from  4500\AA\  to 6600\AA, 
plus H$\alpha$, over 28~nights in 2005.  The lightcurves show intrinsic stellar variability
due to starspots with an amplitude in the blue band of 0.051~magnitudes and a period of 4.847~days.
In addition, three large flares were detected in the data which all occur near the minimum brightness 
of the star.  We remove the intrinsic stellar variability and combine the lightcurves of all the 
filters in order to search for transits by possible planetary companions orbiting in the plane
of the nearly edge-on debris disk.  The combined final lightcurve has a sampling of $0.35$~minutes 
and a standard deviation of 6.8~millimags (mmag).  We performed Monte Carlo simulations by adding 
fake transits to the observed lightcurve and find with 95\% significance that there are no Jupiter 
mass planets orbiting in the plane of the debris disk on circular orbits with  periods, P~$\le 5$~days.  
In addition, there are no young Neptune-like planets (with radii 2.5$\times$ smaller than the young Jupiter) 
on circular orbits with periods, P~$\le 3$~days.
\end{abstract}

\begin{keywords}
stars:individual:AU Mic -- stars:late-type -- circumstellar matter -- stars:pre-main-sequence --
planetary systems
\end{keywords}

\section{Introduction}
\label{sec:intro}

Observing planets around other stars and in various phases of their
evolution is essential to understanding global properties of planetary systems and
necessary for placing our own solar system in the context of planet formation theory.
The detection of a young planet of known age in which the mass and radius can be
measured (i.e.\ a transiting planet) would provide an unprecedented empirical test of
extra-solar planet models and give information on the timescale of planetary formation.
Further, the detection of a planet orbiting a debris disk star would allow investigation 
of the link between the planet forming disk and the planet itself.
We have initiated a project designed to address these questions by searching
for transiting planets around AU Microscopii (AU Mic, GJ~803). 

AU Mic is a young ($\sim 8-20$~Myr, \citet{Barrado}), nearby ($9.94\pm0.13$~pc, \citet{Perryman}) 
active M-dwarf (M1Ve) star in the $\beta$~Pictoris moving group
which is surrounded by a nearly edge-on circumstellar debris disk \citep{Kalasa,Krist}.  
Observations of substructure in the disk suggest the presence of 
planetary-mass bodies.  Such a companion which is close to the star and orbiting in the 
plane of the disk would transit, causing a dip in brightness.  Thus, AU~Mic is an excellent
target to search for transiting planets in the early stages of formation.

Our photometric monitoring campaign that is designed to detect the
signature of a transiting companion orbiting AU Mic is presented in this paper.  
The motivation for the project is discussed in \S~\ref{sec:target}, 
%including a review of the literature on the debris disk.  
In \S~\ref{sec:obs}, the observing program, the data processing, and the determination of
differential photometry are explained.  
The results of the transit search and the parameter space in which
we are sensisitive to planets are described in \S~\ref{sec:transit}.  In \S~\ref{sec:variability},
we briefly discuss the intrinsic variability of the star, and conclusions and future work can
be found in \S~\ref{sec:concl}.

\section{AU Mic and the Debris Disk}
\label{sec:target}

%At the age of $\sim$5-10~Myr, primordial circumstellar star-forming disks 
%are thought to dissipate \citep{Strom93,Haisch01b} leaving planets and 
%planetesimals in the circumstellar environment.  Collisions between large
%planetesimals generate dust and  lead to the formation of a second generation debris 
%disk whose thermal emission is observed in mid-IR to submillimeter wavelengths.  
%The small dust grains in the disk must be constantly replenished
%by collisions from larger planetessimals whose existence is inferred, 
%but not directly observed (**ref**).
%Debris disks are typically optically thin and gas poor and contain cool
%dust beyond tens of AU from the central star.  They have old enough ages such that
%primordial dust can be removed from the system through 
%radiation pressure, Poynting-Robertson drag, or stellar winds
%and larger planetesimals have time to undergo collisions. 

Direct observations of debris disks show small scale structure, including
clumps and rings of dust and gaps clear of dust 
\citep[e.g.][]{Weinberger, Greaves, Clampin, Holland}.  
Orbiting planets could give rise to 
such substructure as planets sweep up material and gravitationally influence the dust.  
A large planet would produce inner gaps cleared of dust, warp asymmetries and clumps of dust 
at mean motion resonances \citep{WyattDent,Wyatt99,QuillThorn} that are observed in debris disks.  

IRAS first detected an excess of 60$\mu$m flux above
purely photospheric emission around AU~Mic, which was interpreted as 
circumstellar dust.  
\citet{Liua} and \citet{Chen} used broadband measurements between 25$\mu$m and 850$\mu$m 
to confirm the 
thermal emission from dust around the star.  A single temperature modified 
blackbody with temperature T = $40\pm 2$K and spectral index $\beta = 0.8$\citep{Liua} 
provides a good fit to the observations.
In addition, recent UV measurements of molecular hydrogen absorption lines
place stringent constraints on the gas-dust ratio of less than 6:1 \citep{Roberge},
indicating the disk is gas poor.  Thus, at an age of $\sim$8-20~Myr
most of the gas has been accreted onto the star or been removed from the system, and
the star no longer contains a primordial, gas-rich star/planet-forming disk.  

\subsection{AU~Mic Disk Structure}

The proximity of AU~Mic has allowed the debris disk to be imaged in
scattered light in both optical 
and IR bands, within $\sim 10-210$~AU from the star, and at resolutions 
as high as 0.4~AU \citep{Kalasa,Krist,Liub,Metchev,Masc}.
Very similar spatial features are observed independently in all the datasets,
suggesting the debris disk is inclined nearly edge-on, is
cleared of dust in the inner regions, and contains smallscale asymmetric 
clumps and gaps of material.

HST+ACS images show an edge-on disk
with an inclination of $< 1^{\circ}$ from the line-of-sight within
50~AU of the star \citep{Krist}.  
\citet{Liub} obtained images with Keck AO which resolved small scale 
structures in the disk in the region from 15-80~AU.    
These authors find asymmetries in the disk that cannot be explained by dust scattering
properties and thus are assumed to be structural.  The disk midplanes are unequal
in length and radially confined bright and dark regions exist in both halves of the disk.

The cool dust temperature obtained through fitting the broad band SED \citep{Liua}
indicates a lack of warm dust in the inner disk near the star.
According to this simple model, the 40K dust temperature translates 
into an evacuated area within 17 AU of the star.  
However, subsequent radiative transfer modelling of the scattered light finds
smaller values of the disk inner radius. An evacuated area within 12~AU from the star
is derived by \citet{Krist} through modelling of the optical scattered light.
\citet{Metchev} combine their high resolution H-band AO data with the existing optical
data and broadband flux measurements to model the radiative transfer through the dust.
The authors find the inner radius of the disk to be $\le 10$~AU.
Due to the bright star, the inner edge
of the disk is not directly observed in any current dataset.

As observations of the debris disk surrounding AU~Mic suggest
the presence of large bodies, several groups have
recently attempted direct detections of thermal emission from  
large planets at distances of 10-60~AU from AU~Mic using AO imaging \citep{Metchev,Masc}.
However, no $>$~1 Jupiter mass (M$_J$) planets were detected at
distances $>$~20~AU from the star and no 5~M$_J$ planets were observed beyond 10~AU.

\subsection{Possible Transiting Planets}
\label{sec:possibletransplanets}

Our group has taken a complementary approach to the search for planets around AU Mic by
targeting the region $<$0.25~AU from the star which is inaccessible with AO technology.  
Due to the edge-on aspect of the 
debris disk, close-in planets orbiting in the plane of the disk will transit the star.
Adopting a $1^{\circ}$ inclination for the disk and assuming a potential planet would
orbit in the plane of the disk, we estimate the maximum period and 
separation at which orbiting planets will undergo transits.  

\begin{figure}
 \epsfig{file=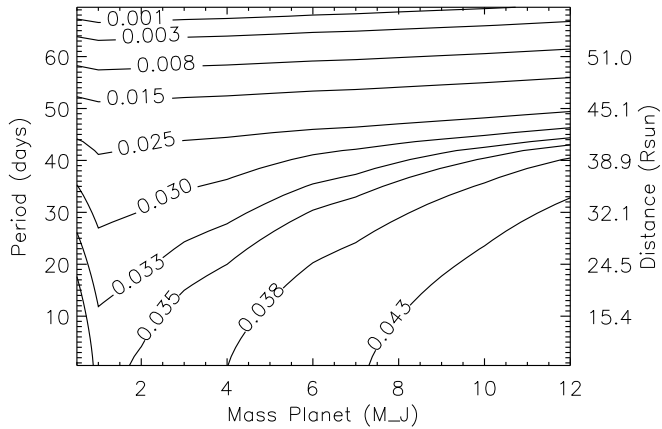,width=3.5truein}
 \caption{Contour plot of the maximum transit depth (mag) as a function of planet
mass and orbital period for potential planets orbiting AU~Mic.  We assume the orbital
plane of the planet is inclined $1^{\circ}$ with the debris disk and adopt properties
for AU~Mic for a 12~Myr, 0.5~M$_{\odot}$ star from \citet{Baraffe98} and planet properties
with masses from  0.5 -- 12~$M_{J}$ from \citet{Baraffe02}.
}
 \label{fig:transdepth}
\end{figure}

Figure~\ref{fig:transdepth}  
shows the maximum transit depth as a function of planet mass and orbital period.
We take the radius for AU~Mic, R=$0.85$~R$_{\odot}$, from the stellar evolution models of \citet{Baraffe98} for
a 12~Myr, 0.5~M$_{\odot}$ star and the planet properties from \citet{Baraffe02} for the same age.
We use the analytic transit model of \citet{MandelAgol}, which includes limb darkening, to derive
the maximum transit depth.  
The theoretical planet radii do not vary much for planets ranging from 0.5 -- 12~$M_{J}$,
thus according to our simple estimates, such planets on circular orbits with 
periods as large as $\sim 70$~days (orbital radius of 55~R$_{\odot}$) will transit
the star.  Planets on $\sim 40$~day orbits will under go full, rather than grazing,
transits causing dips in the lightcure of $> 30$~mmag.  
%We initiated a photometry monitoring campaign in search of transits
%by planetary companions orbiting with periods less than $\sim 70$~days.

\section{Observations}
\label{sec:obs}

\subsection{Observing Program}

We monitored AU~Mic between July 17 and August 14, 2005 with the CTIO-1m telescope and 
Y4K-Cam camera.  The detector consists of a 4Kx4K array of $15\mu$ pixels 
placed at Cassegrain focus giving a $0.3^{\prime\prime}$/pixel platescale. Thus the entire 
array projects to a $20^{\prime}\times20^{\prime}$ FOV.  The observed signal is fed into 
four amplifiers causing the raw images to have a quandrant effect with the readnoise 
between 11-12~$e^{-}$ and gain of 1.45-1.52~$e^{-}$/ADU, depending on the amplifier.
The detector has a readout time of 51 seconds and a
71k-electron well depth before non-linearity sets in. This converts to a
saturation of 40,000 counts/pixel in $1\times 1$ binning mode.  
However, with $2 \times 2$ binning, the readout time is reduced to 
16~seconds, and the peak binned pixel value is limited by digital 
saturation to 65,535 counts.

AU~Mic and the surrounding field were observed in four optical filters of medium width.
These include a custom filter (4520/200\AA), Stromgren-y (5497/241\AA), 
and H$\alpha$-off (6600/75\AA) for continuum monitoring and  H$\alpha$ (6563/75\AA) 
for activity monitoring.  The Stromgren-y, H$\alpha$, and H$\alpha$-off filters, were available 
at the CTIO-1m for use with the Y4K-Cam.  
An image quality 3-cavity custom filter with central wavelength
$\lambda_c$=4520\AA\ and width=200\AA\ was ordered from Custom Scientific.  The filter
is $4\times 4\times 0.3$~inches and is coated with anti-reflection material.
%Out of band blocking outside the relevant region was specified to reduce leakage from the 
%bright red part of the stellar spectrum.  

The narrow continuum filters allow us to isolate specific
regions of the continuum free of highly variable chromospheric emission lines
for optimal transit detection.  The filters avoid the Hydrogen Balmer lines and the Ca II H \& K lines
associated with chromospheric emission, as well as the strong
He features at 4026\AA, 4686\AA, and 5875\AA, and the Na I doublet (5890\AA\ and 5896\AA)
where emission peaks have been observed in absorption line cores of active M-dwarfs.
In practice, the wavelength region covered by the Stromgren-y filter is shared by 
neutral metal lines (e.g.\ Fe~I, Mg~I, Ti~I), and active M-dwarf stars are known to show
very faint emission in these features during flares (e.g.\ \citet{Fuhr,Paulson}).  

The filters provide a 
wide spectral coverage to aid in distinguishing non-gray starspot variability
from gray transits.  In addition, the narrow filters allow us to take
longer exposures that do not saturate the bright target star and that 
are less sensitive to instrumental systematics.
Finally, we monitored in H$\alpha$ to identify residual variability 
in our continuum lightcurves caused by chromospheric activity.  

Throughout each oberving night, we monitored in all four filters alternating 
between H$\alpha$ and one of the continuum filters, systematically cycling through
the continuum filters.  We adopted $2\times 2$ binning to obtain a faster readout 
time on the detector (hereafter, {\it pixel} refers to the binned $2 \times 2$ CCD pixel).  
Our observing program was designed to place the target on exactly the same 
detector pixels in order to minimize inaccuracies due to flat-fielding.  In reality, the 
position of AU~Mic varied within $\sim 5$~pixels from the chosen position.  
Exposure times were chosen to maximize the flux in the target star and nearby 
reference stars while keeping the peak pixel value in AU~Mic below $\sim 60,000$ counts. 
We defocused the telescope to avoid saturating AU~Mic while taking longer
exposures to build up signal in the fainter reference stars.  
We monitored during non-photometric weather and changed the exposure time 
continuously based on the photometric transparency.  Thus, exposure times were varied 
between $3-40$~seconds in all bands.  In this way, a 
median sampling rate of 0.8~minutes was obtained for H$\alpha$ and $\sim 2.5$~minutes
for all of the continuum filters.
We observed AU~Mic for 6-10 hours per night on 19 nights during that time.  Due to poor weather, 
we obtained only 1-2 hours of data on 5 nights and completely lost an additional 5 nights.

\subsection{Processing the Images}

Flat field and bias calibration frames necessary for processing the images
were obtained during each observing night.  We took 2-D bias frames approximately every
few hours in addition to sets of biases at the beginning and end of each night.
At least 10 dome flats were observed per night in all four filters, and
twilight flats (3-4 per filter) were obtained when the weather was clear.
%The twilight flats provide a good match to the large scale flat-field
%structure in the images which does not change significantly over the course
%of the run. The nightly dome flats allow us to apply a high signal-to-noise
%correction for small scale features (e.g.\ dust "donuts") which vary from
%night to night.

The images were processed in a standard way using routines written by L.\ Hebb in the IDL
programming language.  Before performing any processing tasks, we checked for bad frames.  
Images in which the peak pixel value in
AU~Mic is equal to 65536 are saturated and those where the peak pixel value is less than 2000 counts
have too low transparency to obtain useful magnitude measurements of the reference stars.
There were typically $\sim$10 out of $\sim$1200 such images on a full 
night of observing which were removed from our processing list.

Each of the four amplifiers was processed independently.  All object and calibration
frames were first overscan corrected (by subtracting a line-by-line median overscan value)
and then trimmed.  We created nightly stacked bias images by average combining all bias frames observed each night.  
However, we noticed a smallscale 'herringbone' pattern in the bias frames which varied over the course
of an observing night.    
This has subsequently been observed by other groups (see http://www.lowell.edu/users/massey/obins/bias.html).  
The amplitude of the variability is at the level of $\pm 10$ counts/pixel, and it persists in our data.
This corresponds to a maximum of 0.1\% of the typical 
AU~Mic flux and 0.3\% of the combined flux of all the reference stars.  
Therefore, it contributes to the noise in the resulting lightcurves
at the milli-mag level.

The stacked bias frames were subtracted from all object and flat field images.  Average combined
nightly dome flats were created in each filter, and each night, the object frames and any twilight 
flats that were obtained were divided by the stacked dome flat.
After applying the dome flat correction, there is still residual large scale flat-field structure
in the images which is stable over the course of the observing run.  
We tested the application of an additional illumination correction creating
a stacked dome-flat corrected twilight flat, and divided this image by the object images.
We applied this correction to several nights of data, but it
did not provide any improvement to the resulting differential photometry.
This is likely because we chose reference stars within $5^{\prime}$ of the target star over
only a part of the detector in which the dome flats were a good match to the flat-field structure.
As it did not improve the photometry, we did not apply this illumination correction to the data.

There is no apparent fringing structure in any of our images which span the wavelength range
from 4500-6600~\AA, including the continuum bands and H$\alpha$.  
Thus, we do not apply a fringe correction.  In addition, the exposure
times are short and the dark current is negligible.  We do not apply a dark current correction
as its application would only add noise to our lightcurves.  
We obtained approximately 3300 observations of AU~Mic in each of the continuum filters
and $\sim$9800 observations in H$\alpha$.  

\subsection{Generating Lightcurves}
\label{sec:dphot}

\subsubsection{Photometry}
After the instrumental signatures were removed, 
source detection and aperture photometry were performed on all science frames 
using the CASU catalogue extraction software \citep{IrwinLewis}.
The software has been compared with  SExtractor
(http://www.ast.cam.ac.uk/$\sim$wfcam/docs/reports/simul/) and found to be very
similar in the completeness, astrometry and photometry tests. 
The source detection algorithm defines an object as a set of contiguous pixels 
above a defined detection threshold.  The routine requires as input the detection
limit in units of background $\sigma$ and the minimum number of connected pixels 
above that threshold which define an object.  We set a detection threshold  
of 3 $\sigma$ and a minimum source size of 15 pixels, so that AU~Mic, the brightest
star in the field, is easily detected in all the images.  Potential comparison
stars up to $\sim 7$~magnitudes fainter are also detected.

Aperture photometry with circular apertures was performed on the 
detected objects in all frames.  The aperture size
affects the precision of the differential photometry, thus the optimum
aperture radius was chosen through empirical testing of several different sizes.
An aperture which is too small will be suceptible to centering errors and can be affected by 
the pixelization of the detector, whereas an aperture which is too large will begin
to include noise associated with sky pixels without adding much additional signal.
Our experience with aperture differential photometry suggests that for bright
stars, apertures larger than the median seeing produce lightcurves with lower rms.

The median seeing of our observations is $\sim 1.5^{\prime\prime}$ which corresponds to
$\sim 2.6$~pixels ($\sim 1.5^{\prime\prime}$).  Circular apertures with radii of  
2.8, 4, and 8~pixels were tested on a 
single night of data.  The same general features are present in the lightcurves
generated with all the tested apertures, including the intrinsic stellar
variability, but the 4~pixel ($2.4^{\prime\prime}$) aperture produced the AU~Mic lightcurve with the 
smallest scatter.  Thus, we obtained instrumental magnitude measurements for all
detected objects on all frames by summing the background subtracted flux 
contained in the 4~pixel aperture.  
%The value of the instrumental
%magnitudes are determined by the apparent magnitude of the star, the overall
%transmission through the atmosphere and the telescope, and the atmospheric extinction.

\subsubsection{Differential Photometry}

To derive a differential photometry lightcurve for AU~Mic from the instrumental photometry,
we first correct for atmospheric extinction
using magnitude measurements obtained on photometric nights for nearby, bright 
comparison stars.  In each filter, we apply a linear least squares fit 
to the instrumental magnitudes as a function of airmass.  
The derived first order extinction coefficients
are as follows: 0.21 for the 4520/200 filter, 0.14 for the 5497/245 filter, and 
0.083 for the 6600/75 and H$\alpha$ filters.  We apply these extinction corrections to 
all instrumental magnitude measurements.  We note that this linear extinction correction 
is not strictly necessary 
since it is applied to all stars equally, and thus is removed when the differential magnitude is
calculated.  However, we correct for this well understood source of flux attenuation to derive
an instrumental lightcurve which is dominated by the fluctuating atmospheric transparency,
especially as caused by thin cirrus and other clouds

Next, we remove points from the lightcurves which were obtained under very poor
observing conditions ($> 1.0$~mags below the magnitude obtained under photometric conditions).  
In addition, we remove a handful of measurements obtained with exposure times $< 3$s which are
adversely affected by the finite shutter opening time.
Finally, we derive the differential photometry lightcurve for AU~Mic using
the measurements of 13 comparison stars within $\sim 5^{\prime}$ and
up to $\sim 5$~magnitudes fainter (in the 4520/200 band) than AU~Mic.  
 
\begin{figure}
 \epsfig{file=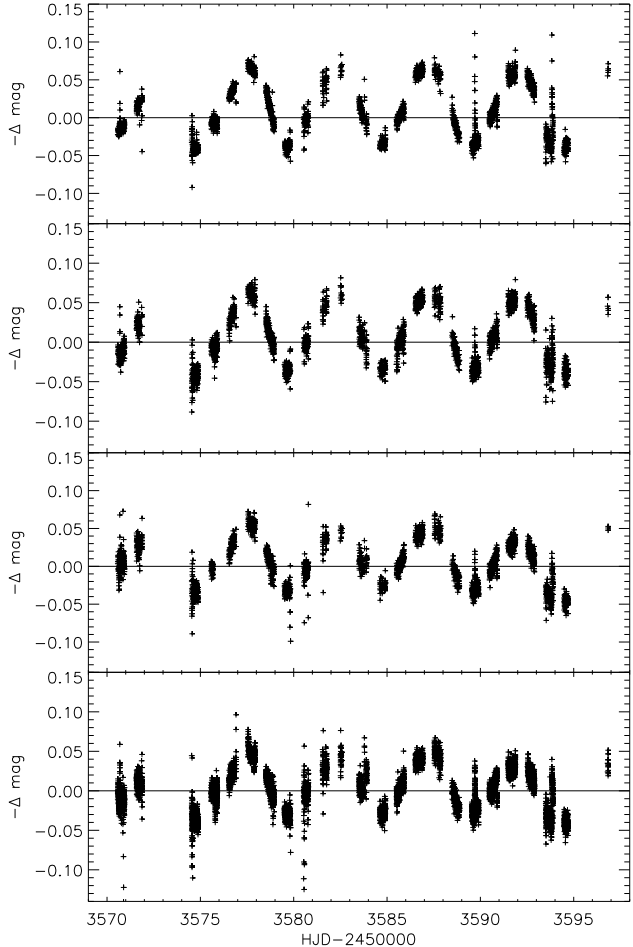,width=5.0truein,angle=90}
 \caption{AU~Mic differential photometry lightcurve in the four filters: F4520 (top), F5497 (middle-top),
F6600(middle-bottom), H$\alpha$ (bottom).  Starspot
  variability and flaring are present in the data.}
 \label{fig:dphot}
\end{figure}

Then we create a {\it super} comparison star lightcurve by averaging the
instrumental magnitude lightcurves for the 13~individual stars 
(after subtracting the median value from each star).  Initially, the 13~reference
stars are weighted equally.  Differential photometry lightcurves are then
obtained for the individual comparison star lightcurves 
by subtracting the {\it super} lightcurve.  Each comparison star is then
assigned a weight value equal to the inverse variance of its differential photometry
lightcurve normalized so that the weights of all 13~stars sum to one.  The {\it super}
comparison star lightcurve is then re-created, this time combining the individual
lightcurves using a weighted average.  The new {\it super} comparison star lightcurve
is then subtracted from the instrumental AU~Mic lightcurve.
The resulting differential photometry lightcurves for AU~Mic are shown in 
Figure~\ref{fig:dphot} for the four filters.  Since AU~Mic is significantly
brighter than the nearby reference stars used to derive the differential photometry, the lightcurve
precision is dominated by photon Poisson noise in the reference star magnitude measurements.  
We have defined an error for each point on the lightcurve by taking the weighted
standard deviation of the 13 comparison star objects used to derive that point on the
AU~Mic lightcurve.  

\subsubsection{Removing Intrinsic Variability}

Figure ~\ref{fig:dphot} shows the intrinsic stellar variability previously
observed in AU~Mic \citep{Rodono}.  We observe flare activity and sinusoidal variability 
indicative of starspots on the stellar surface.
Both sources of variability are linked to magnetic activity and are typical 
for a young, active M-dwarf star, like AU~Mic.  
The instrinsic variability of AU~Mic is discussed briefly in \S~\ref{sec:variability}.
However, the variability must be removed in order to search for the signature of transiting 
planets in the lightcurve.

Both the shape and the timescale of these sources of variability differ significantly from 
that of a transit, so they do not induce contamination in the transit search.  
During our observing campaign, three large flares occured on AU~Mic which were detected
in all four filters.  During the flares, the magnitude of AU~Mic increased sharply
within minutes and then decayed slowly to its original value over an hour to a few hours. 
We simply remove the regions of the AU~Mic lightcurves which include the three large flares.
Table~\ref{tab:flarereg} gives the range of Heliocentric Julian Dates (HJDs) which are used
to exclude points from the lightcurve near the time of the flares.

\begin{table}
\caption{HJD of Flare Cutout Regions }
\label{tab:flarereg}
\centerline{
\begin{tabular}{ccc}
\hline\\[-5pt]
Flare & Beginning & End \\[5pt]
\hline\\[-5pt]
Flare 1         & 2453570.68  & 2453570.73   \\
Flare 2         & 2453589.69  & 2453589.77   \\
Flare 3         & 2453593.82  & 2453593.90    \\
\end{tabular}
}
\end{table}

The quasi-sinusoidal, starspot variability occurs on timescales of $\sim 5$~days and varies in
amplitude for the different filters.  The Hipparcos period for this object is
4.8902~days, however we find the variation occurs on a 4.847~day timescale 
(see \S\ref{sec:variability}).  To remove the modulation, we experimented with
fitting a truncated fourier series to the phase-folded lightcurves in each filter
as well as applying linear least-squares fits to the individual nights of data.
We note that neither approach is a physical interpretation of the data or an attempt to model
the starspots.  Both techniques produce corrected lightcurves with similar noise properties, 
thus, for this analysis, we subtract a linear fit from each night of data 
to produce the AU~Mic lightcurve for each filter which is removed of intrinisic variability.
Table~\ref{tab:lcinfo} gives the number of measurements in the corrected lightcurve
for each filter, as well as the median sampling, the rms (weighted by the error bars) and the
correlated noise on a 2-hour timescale, the typical planetary transit duration.
To measure the correlated noise, 
we calculate the rms of the lightcurve where each point is replaced by the average
of the points in a 2-hour window around that point (accounting for edge effects) \citep{Pont06}. 
The rms of the smoothed lightcurve is 1.7~mmag which is higher than what is expected 
if the data were only white noise.

\begin{table}
\caption{Time Series Information} 
\label{tab:lcinfo}
\centerline{
\begin{tabular}{ccccc}
\hline\\[-5pt]
Filter & Number & Median   & Lightcurve & Correlated      \\
       & Points & Sampling & RMS        &   Noise         \\
       &        & (min)    & (mmag)      &    (mmag)        \\
\hline\\[-5pt]
F4520         & 3089   &  2.46 & 5.3 & 1.8   \\
F5497         & 2960   &  2.46 & 7.3 & 1.8   \\
F6600         & 2800   &  2.46 & 7.2 & 2.1   \\
F6563         & 8737   &  0.70 & 7.6 & 2.3   \\
Combined      & 17593  &  0.35 & 6.8 & 1.7    \\
\end{tabular}
}
\end{table}

\begin{figure*}
 \epsfig{file=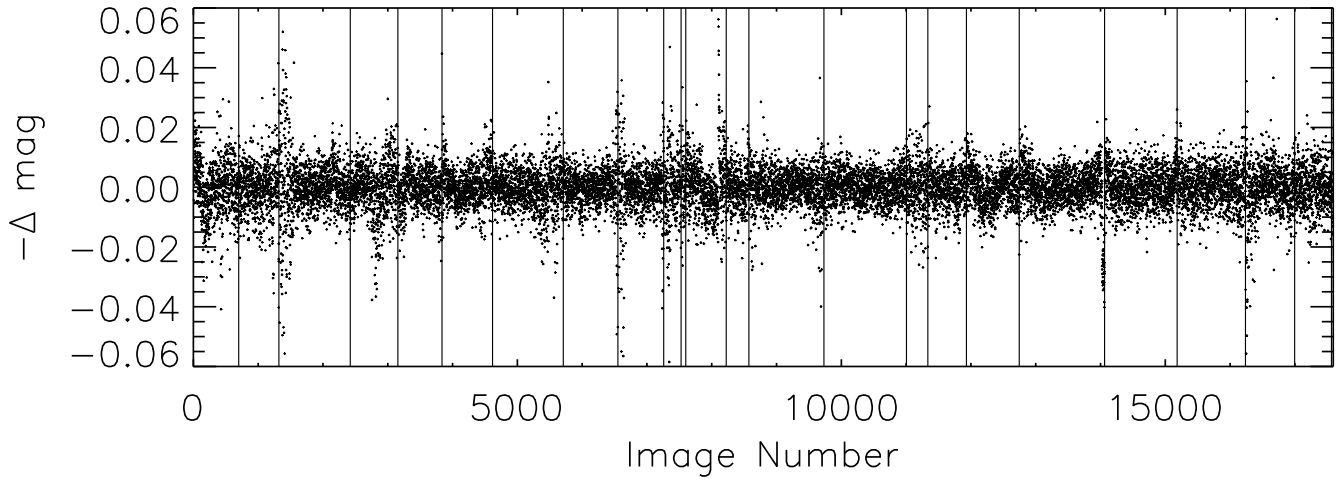,width=3truein,angle=90}
 \caption{Combined, intrinsic variability-corrected, differential photometry 
lightcurve of AU~Mic plotted versus image number.
 The data contain differential magnitudes for all four filters.  
They are removed of flares and corrected
 for the rotationally modulated starspot brightness. 
 The vertical lines are placed at the positions of the night breaks. }
 \label{fig:finallc}
\end{figure*}

Since the transit signals for which we are searching will produce
dips in brightness of the same depth in all four filters, we combine the four
corrected lightcurves into a combined final AU~Mic lightcurve, shown in
Figure~\ref{fig:finallc}.  The combined, intrinsic variability-corrected lightcurve 
is input into the periodic transit 
searching algorithm.  The lightcurve contains 17593 points, has a median
sampling of 0.35~minutes, and has an rms of 6.8~mmag (see  Table~\ref{tab:lcinfo}).

\section{Search for Transiting Planets}
\label{sec:transit}

\subsection{Systematic Search for Periodic Transits: Applying the Box-Least-Squares Algorithm}

The final, combined, intrinsic variability-corrected lightcurve was searched for features
that could have been caused by a transiting planet.  We performed a systematic search for periodic,
square-shaped dips in brightness using the Box-Least-Squares algorithm of \citet{Kovacs}.
We apply the algorithm to the final lightcurve, testing periods between $0.5-15$~days.
The short period limit is set
by the Roche radius of a Jupiter mass planet around AU~Mic, and the long period
limit is approximately one half the duration of our monitoring campaign.  Known short period,
hot Jupiter planets have transit durations of a few hours, thus we search for box-shaped dips
with widths corresponding to planet durations from $\sim 1-5$~hours.

We employ a detection statistic, 
which identifies the signal-to-noise ratio, S/N, of the strongest dip in brightness as
a function of trial period.  The signal, S, is the depth of the candidate dip, and the noise,
N, accounts for both random and correlated noise \citep{Pont06}
as both contribute significantly to the noise in our final lightcurve.    The S/N is given by:
$${\rm S/N} =\frac{depth}{\sqrt{\sigma^2/N_{intr}+\sigma_{red}^{2}/N_{tr}^{2}  }}$$
where $depth$ is the average depth of the transit-like event, $\sigma$ is the average
photometric error of the in-transit points, $N_{intr}$ is the number of in-transit
points, $\sigma_{red}$ is the correlated noise on the typical transit
timescale (2 hours) and $N_{tr}$ is the number of detected transits.  The resulting
periodogram (S/N versus trial period) was then examined for significant peaks which would indicate periodic box-shaped
dips in the lightcurve.  We set a threshold of S/N~$=8.8$ to define a significant detection. 

%\subsubsection{Setting the Detection Threshold}

The detection threshold of S/N~$=8.8$ was derived from analysing a set of simulated lightcurves
which contained only noise (no fake transits).  We generated a set of 400 fake 
lightcurves with the same window function and noise properties of the observed lightcurve.
This allows structure in the periodogram which arises due to the
window function to mimic that of the observed lightcurve.  The delta magnitude values are
different for each simulated lightcurve and consist of the sum of a red noise component
and a white noise component.  The noise in each simulated lightcurve is randomly determined
using the measured values given in Table~\ref{tab:lcinfo}. 

%To create the red noise component of the delta magnitude, we first select random values
%from a Gaussian distribution with a mean of zero and a sigma of 1.7~mmag, the level of
%the correlated noise on a 2-hour transit timescale in the observed data.  The values are paired
%with time values spaced at two hour intervals (with random phase) for the duration
%of the lightcurve.  We employ a spline interpolation to derive red noise values for 
%all time points in the lightcurve between the 2-hour time intervals.  The white noise
%component of the delta magnitude is generated by drawing random points from a Gaussian 
%distribution with a mean of zero.  The width of the random white noise Gaussian
%is set at 6~mmag to make the width of the total distribution of delta magnitudes (including
%red and white noise) in the simulated lightcurves match that of the observed data.  

The 400 simulated lightcurves, including only noise, were searched for transits in the 
same fashion as the observed data.  
We applied the box-fitting algorithm to each lightcurve and measured the peak S/N value
that was detected.  The algorithm typically finds a low level
spurious signal with a depth of $\sim$2~mmag and a S/N~$\sim 6$ in the simulated data.  
For the ensemble of simulated lightcurves, 
the distribution of output S/N values has a mean of 6.1 and a sigma of 0.9.  99.7\% (3$\sigma$) of 
the simulated lightcurves produce a peak S/N$ < 8.8$, so we adopt this value as the detection
threshold to define significant periodic transits in the real data.

\subsection{Detection and Analysis of Event: Is it Caused by a Transiting Planet?}
\label{sec:eventanal}

\begin{figure}
 \epsfig{file=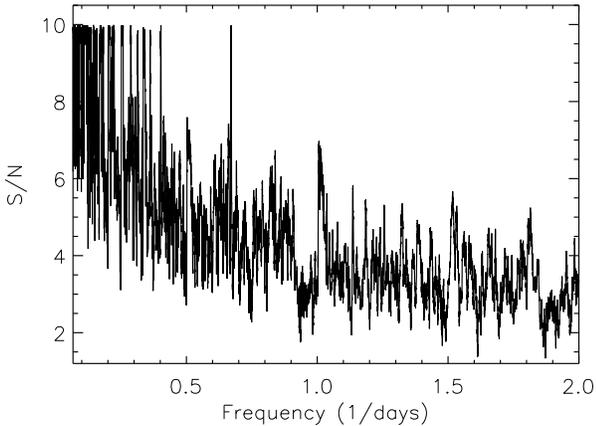,width=3.5truein,angle=0}
 \caption{Periodogram resulting from the box-fitting algorithm. }
 \label{fig:transpdgrm}
\end{figure}

The periodogram resulting from the periodic transit search on the intrisic variability corrected
AU~Mic lightcurve is shown in Figure~\ref{fig:transpdgrm}.  There are many peaks 
in the periodogram above the significant detection level of S/N=8.8.  From this, we infer
that a significant box-shaped dip in brightness is present, however a unique period
is not identified for it.  Only one dip is apparent in a close visual inspection of the
lightcurve, and the multiple peaks in the periodogram are likely due to the single event 
aliasing with the window function.

\begin{figure}
 \epsfig{file=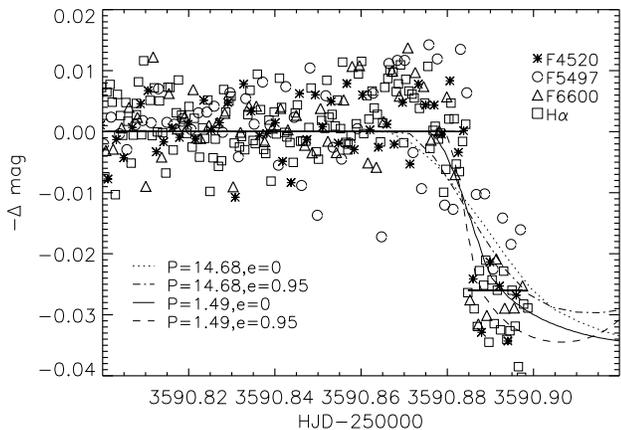,width=3.5truein,angle=0}
 \caption{Lightcurve of AU~Mic zoomed in on the detected dip in brightness.
The data from the four filters are shown as different symbols (F4520,asterisks;
F5497, circles; F6600, triangles; H$\alpha$,squares).  
Four transit model lightcurves from \citet{MandelAgol} are overplotted.
The model lightcurves correspond to planets with orbital periods and eccentricities of:
P=1.49~days, e=0.0 (solid), P=1.49~days, e=0.95 (dashed), 
P=14.48~days, e=0.0 (dotted), and P=14.48~days, e=0.95 (dot-dashed).  
We adopt a stellar radius of  R$=0.85$~R$_{\odot}$ (12~Myr, 0.5~M$_{\odot}$ star \citep{Baraffe98}
and a planet radius of 0.13~R$_{\odot}$ to roughly reproduce the depth of the dip. }
\label{fig:dip}
\end{figure}

The single observed dip in brightness occurs in all four filters during the last 20~minutes 
of night JD~2453590.  It consists of 39 data points, beginning at HJD$=2453590.884766$ 
and continuing until the observing is stopped for the night at HJD$=2453590.898682$.  
The average depth is 26.4~mmag, resulting in a S/N~$=10.0$.
Figure~\ref{fig:dip} shows a blow-up of the lightcurve region which includes the transit-like event.
The dip is clearly shown and is highly significant, however, it is not repeated,
the ingress is very sudden, and the egress is not observed.  
Therefore, it is possible the dip is due to an instrumental, rather than astrophysical cause.

\subsubsection{Considering Instrumental Explanations}

We first examined the lightcurves of the individual reference stars for the
night of JD~2453590.   The reference stars showed no sudden brightness
variations at the end of the night which would have caused the observed dip in 
the differential magnitude of AU~Mic.  We then checked for trends
in the AU~Mic lightcurve with instrumental properties.  We plotted
the data against the peak counts in AU~Mic and its position on the detector,
but found no correlation with the observed dip. On the same night, we checked for a relationship
between the differential magnitudes and their corresponding airmasses, 
exposure times, background levels, and seeing measurements.  Again,
we found no correlation between these instrumental properties and the differential photometry
which could have caused the drop in brightness.  In addition, we tested
several photometric aperture sizes, and all showed the same dip in the lightcurve
with the same depth and timing.  Thus, we identified no systematic instrumental cause for the dip. 
However its nature is still uncertain.  To determine whether the drop in brightness could 
have been caused by a transiting planet, we investigate the type of planet that would be 
required to be consistent with the observed lightcurve.  

\subsubsection{Depth and Period Constraints}

If a transiting planet caused the observed dip in brightness, it would have to be quite large.
The relationship between transit depth and planet radius can be approximated
by $depth \sim 1.3 \times (R_p/R_{s})^2$ \citep{TingSack}.  Given a measured depth of 26.4~mmag
and adopting a radius for AU~Mic of 0.85~R$_{\odot}$ \citep{Baraffe98}, the planet radius
is found to be $R_p \sim 1.1$~Jupiter radii.  Although there are large uncertainties in the stellar
evolution models used to define the stellar radius, and the depth equation is an approximation
which is strictly only valid in the I-band, it is clear that only a planet with a Jupiter-like radius 
could account for the observed dip.

A transiting planet should show multiple dips in brightness occuring at periodic intervals.
This was not observed.  However, due to the window function of the lightcurve, it is possible that additional 
events occured while we were not observing (e.g.\ during daytime or poor weather).  Therefore,
there are only certain periods between 0.5--15~days which are consistent with a transiting
planet as the cause for the observed event.  Peaks in the periodogram identify potential periods,
which all have identical S/N values because they identify the same dip in brightness.
The shortest possible transiting planet periods which 
are still consistent with the data are P=1.492~days, 2.484~days, and 3.870~days.  The number
of such periods increases with orbital period, with the longest one (within our search range) being P=14.680~days.  
Although, we are not sensitive to periodic signals longer than the duration of our observing
campaign, we note that many possible planet periods outside our search range, between  15--70~days 
(see \S~\ref{sec:possibletransplanets}), are also possible.

\subsubsection{Constraints from the Ingress}

The observed dip in brightness is very sharp with little or no aparent ingress.  
To determine whether a planet could cause such a sharp transit, 
we compared the observed dip to model lightcurves from \citet{MandelAgol}.
Four model lightcurves which represent the extremes of the potential planets and which
are consistent with the depth and period constraints defined above 
are overplotted on the data in Figure~\ref{fig:dip}.  Both the planet period
and eccentricity affect the shape and duration of the ingress (if the depth of the
transit and radius of the star are fixed), thus we try two different orbital
periods of P=1.49~days and P=14.68~days, which both correspond to peaks 
in the periodogram.  For both periods, we show the ingress curve of a planet with
an eccentricity of $e=0.0$ and $e=0.95$ to represent the extremes at each period.  
We adopt a stellar radius, R$_{s}=0.85$~R$_{\odot}$, for a 12~Myr, 0.5~M$_{\odot}$ star
\citep{Baraffe98} and a planet radius, R$_{p}$~=0.13~R$_{\odot}$, which is consistent
with a Jupiter mass planet of this age and roughly reproduces the depth of the feature 
in the observed lightcurve.  A quadratic limb darkening law is used with parameters 
found in \citet{Claret} for a star with T$_{eff}=3500$ K and logg$=4.0$.  

The long period planet models (P~$=14.68$~days) shown here are inconsistent with the data
due to their relatively long ingress compared to the sharpness of the observed dip.  
Consequently, planets on longer period orbits would also be unable to explain the data.
The short period planet (P~$=1.49$~days) on a circular orbit ($e=0.0$) is also largely
inconsistent with the observed sharpness of the dip.  The ingress of the short period planet 
on the highly eccentric orbit provides the closest match to the shape of the dip, however it
still does not replicate the sharpness of the observed feature.

In summary, we placed constraints on the properties of a potential orbiting planet that could
have caused the drop in brightness of AU~Mic that was observed on JD~2453590.
The combination of the deep, sharp, non-repeated dip makes it unlikely that
the observed feature, shown in Figure~\ref{fig:dip}, was caused by a transiting planet.
In addition, there is no reason for us to believe the feature is caused by an instrumental problem.
Therefore, we conclude this single event which mimicks a planetary transit is unexplained,
and additional data is required to determine its true nature.

\subsection{Search for Smaller Planets}

In order to search for smaller planets in the data, we first remove the dip in brightness
discussed in \S~\ref{sec:eventanal} by subtracting the average depth of the dip 
(26.4~mmag) from the data points between HJD$=2453590.884766$ and HJD$=2453590.898682$.  
We then re-run the box-fitting algorithm.  In this trial, the highest peak in the periodogram has a 
S/N=$6.4$, and thus is only consistent with noise.  Therefore, we do not detect any significant,
low-amplitude periodic box-shaped dips in brightness.  The implications of this non-detection
are discussed below.  

%  the model lightcurve derived
%from the parameters found in the initial transit search and then re-search for additional
%box-shaped features in the data.   The parameters of the initial transit search produce a model 
%lightcurve in which square-shaped dips in brightness occur periodocially with a period 
%of $P=2.942$~day, a depth of 26.4~mmag, and a duration of 1~hour.  The phase of the model lightcurve
%is such that one of the model transits begins at HJD~$2453590.8845$ to coincide with the observed
%dip in brightness.  Therefore, a constant 26.4~mmag value is subtracted from the observed lightcurve 
%in the region of the detected event.  The remainder of the observed lightcurve is unaltered as 
%the other transits in the model lightcurve coincide with gaps in the data.  

\subsection{Fake Transit Simulations}
\label{sec:faketrans} 

To set limits on the type of planet which could have been detected in our data, 
we generated simulated lightcurves with fake transits added to the observed data
and tried to recover the transit signal.  

In the simulations, we placed a planet in orbit around AU~Mic in the plane of the debris disk
(inclination angle equal to $1^{\circ}$ from the line-of-sight).  We adopted a host star radius of
0.85~R$_{\odot}$ for a 12~Myr, 0.5~M$_{\odot}$ star.
The orbital phase of the simualted planet was chosen randomly from a uniform distribution, and
the orbital period of the planet was chosen randomly from a uniform distribution within our
search range (0.5--15~days).  All simulated planets are assumed to be on circular orbits.
We made no attempt to adopt the distribution of orbital properties
of known planets since there are no observed data to constrain these values at the age of AU~Mic.
Noise-free model lightcurves were created using the
analytic eclipse models of \citet{MandelAgol} which were then added to a version of the
observed data in which the detected dip in brightness described in \S~\ref{sec:eventanal} was subtracted off.
Thus, the simulated lightcurves with fake transits added have the same noise properties
and sampling as the observed lightcurve.
  
The fake transit lightcurves were run through the box-fitting algorithm using the same search parameters as
the observed data (searching for periods between 0.5--15~days and durations of $\sim$~1--5~hours).
The fake transits were considered recovered if the transit signature was detected with a
S/N~$\ge 8.8$ and the derived period was within 5\% of the input orbital period.  Alias periods
which are one half or twice the input period are also considered recoveries.  

\begin{figure}
 \epsfig{file=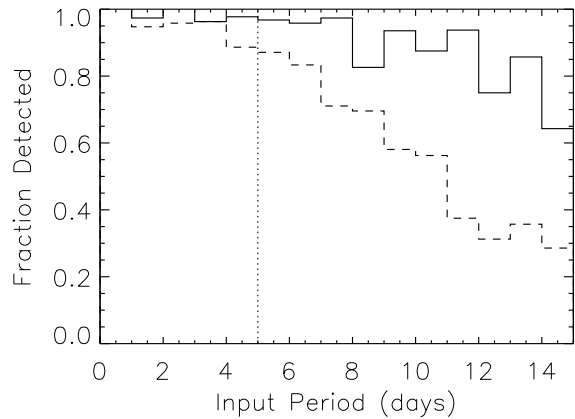,width=3.5truein,angle=0}
 \caption{Fraction of simulated Jupiter-mass planets detected in the fake transit Monte Carlo simulations
as a function of input orbital period.  The solid line shows the fraction detected with a S/N~$> 8.8$, and the dashed line
 shows the recovery fraction with a S/N~$>8.8$ and in which the period found by the 
box-fitting algorithm is within 5\% of the input period (including twice and one-half aliases) (dashed line).
The vertical dotted line corresponds to the 5~day period limit which is discussed in \S~\ref{sec:faketrans}.
}
\label{fig:addtrans1}
\end{figure}

We ran two sets of simulations (with 400 lightcurves each) using two different planet radii.
First, we simulated a $1 M_{J}$ planet in orbit around AU~Mic adopting  radius,
R$_{p}$=0.134~R$_{\odot}$ for a 12~Myr, 1~M$_{J}$ planet from \citet{Baraffe02}.  Such a planet
produces an unmistakable signal in the simulated lightcurves with a depth of $> 30$~mmag.
%For periods, P$\le 15$~days, in 93\% of the cases, the box-fitting algorithm detects at least 
%one transit with S/N~$> 8.8$, and 71\% of the time, the period is also recovered correctly.  
The recovery fraction is a strong function of 
input period, as is shown in Figure~\ref{fig:addtrans1} which plots the fraction of transiting 
Jupiter mass planets detected in the simulations as a function of input orbital period.  If we consider only short period
planets with periods, P$\le 5$~days, 99\% of the sample are significant detections 
with a S/N~$> 8.8$ (solid line in  Figure~\ref{fig:addtrans1}), and for 95\%, 
the period is also correctly recovered (dashed line).
%97\% have at least 2 transits in the lightcurve which are detected with S/N~$> 8.8$ (dot-dashed line), and for 95\%, 

\begin{figure}
 \epsfig{file=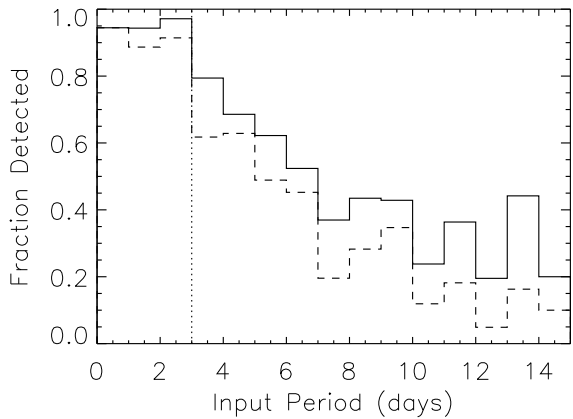,width=3.5truein,angle=0}
 \caption{Results of the fake transit Monte Carlo simulations.  Fraction of simulated Neptune-like planets, 
with a mass, M=0.05~M$_{J}$, and a radius, R$=0.054$~R$_{\odot}$,
which is 2.5 times smaller than the radius of the 12~Myr old Jupiter.  The solid line shows
fraction detected with a S/N~$> 8.8$, and the dashed line shows the recovery fraction
with a S/N~$>8.8$ and in which the period found by the 
box-fitting algorithm is within 5\% of the input period (including twice and one-half aliases) (dashed line).
The vertical dotted line corresponds to the 3~day period limit which is discussed in \S~\ref{sec:faketrans}.
  }
 \label{fig:addtrans2}
\end{figure}

We ran an additional set of simulations using a planet which represents a possible Neptune-like
planet at the age of AU~Mic.  Since the theoretical models do not reach the mass of Neptune
and no young Neptune mass planets have been observed, we simply adopt a mass, 
M~$=0.05$~M$_{\odot}$, and a radius, R$=0.054$~R$_{\odot}$, 
which is 2.5 times smaller than the radius of the 12~Myr old Jupiter.  This type of planet 
produces a transit depth of $\sim 5$~mmag which is smaller than the noise limit of the data.
However, for orbital periods, P$\le 3$~days, 95\% of the simulated transit lightcurves are detected
with a S/N~$> 8.8$.  In 92\% of the cases, the period is also recovered.  Figure~\ref{fig:addtrans2}
shows a plot of the fraction of transiting Neptune like planets detected in the simulations as a
function of orbital period.
%All of which have two or more transits in the lightcurve. 

\subsection{Interpretations}
\label{sec:results}

We interpret the results of the transit search on the observed lightcurve in the context
of the fake transit simulations.  Due to the lack of a convincing planetary transit
detection in our observed lightcurve, we find, with 95\% confidence, that there are no
Jupiter mass planets on circular orbits around AU~Mic with periods $\le 5$~days. 
We find at the 92\% confidence level that there are no young planets with smaller 
radii (Neptune-like) orbiting AU~Mic on circular orbits with periods $\le 3$~days.

It is important to note that the Monte Carlo simulations described above are based 
on theoretical stellar 
evolution models which are uncalibrated and untested by observations for low-mass stars and 
planets at the young age of AU~Mic \citep{SuzMonitor}.  In addition, early M-dwarfs are 
rapidly evolving in radius between 10-20~Myr, therefore, uncertainties in the age of AU~Mic 
will produce uncertainties in the results of our simulations.

\begin{figure}
 \epsfig{file=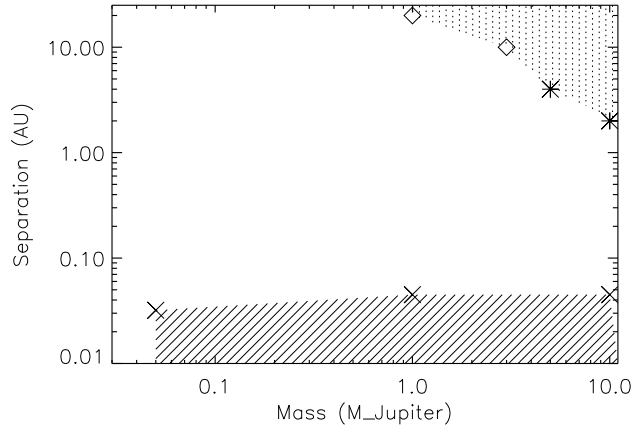,width=3.5truein,angle=0}
 \caption{Region of planet parameter space as a function of planet mass and orbital separation 
showing the constraints on the existence of planets around AU~Mic given by adaptive optics 
from \citet[diamonds]{Metchev}
and \citet[asteriks]{Masc} and time series photometry (this work, crosses).
The hashed areas show regions of the mass/separation parameter space in which the existence of a planet is ruled 
out with high significance.  
  }
 \label{fig:pspace}
\end{figure}

We combine our results with the constraints placed by adaptive optics imaging \citep{Metchev,Masc} 
on the existence of planets orbiting in the outer parts of the AU~Mic disk.  In Figure~\ref{fig:pspace}, 
we show the region of parameter space as a function of planet mass and orbital separation 
in which an existing planet would have been detected.  The existence of a planet in the hashed region
of the diagram is ruled out with high significance due to the lack of a secure planet detection in either
the adaptive optics or photometric monitoring dataset.  There is still a large region of parameter space
left to be explored.  The search for transiting planets can potentially explore beyond the existing
constraints out to 0.25~AU. Radial velocity measurements should be able to detect planets in the system
out to several AU, however, the stellar activity will reduce the detection sensitivity of this method.

\section{Intrinsic Stellar Variability}
\label{sec:variability}

In addition to the planet search, the high cadence, multi-band photometric dataset described in this paper
provides information about the intrinsic variability of AU~Mic.  We briefly report on the findings,
but save a more in depth discussion, analysis, and modelling for a future paper.

\subsection{Starspot variations}
\label{sec:starspot}

The stellar variability due to starspots is quasi-sinusoidal and periodic on the timescale of the stellar rotation
period.  We measure the periodocity in the lightcurve using a 
Lomb-Scargle periodogram \citep{Scarg} combining data from all the filters.  
The main peak in the periodogram is highly significant and occurs at 4.847~days.
Our data covers approximately
six rotations of the star, however with a period very close to 5 days, the phase coverage is incomplete.
The period we derive is closer to the 
4.854~day period found by \citet{Torres} than to the Hipparcos period of 4.8902~days.

%\begin{table}
%\caption{Amplitude \& Epoch of Variability}
%\label{tab:ampvar}
%\centerline{
%\begin{tabular}{ccc}
%\hline\\[-5pt]
%Filter    & Amplitude & Epoch   \\ 
%          & (mag)     & (days)  \\ 
%\hline\\[-5pt]
%F4520             &  0.051 &  2453571.3603  \\
%F5497             &  0.047 &  2453571.3357\\
%F6600             &  0.039 &  2453571.3202\\
%H$\alpha$         &  0.035 &  2453571.3143\\
%\end{tabular}
%}
%\end{table}

\begin{figure}
 \epsfig{file=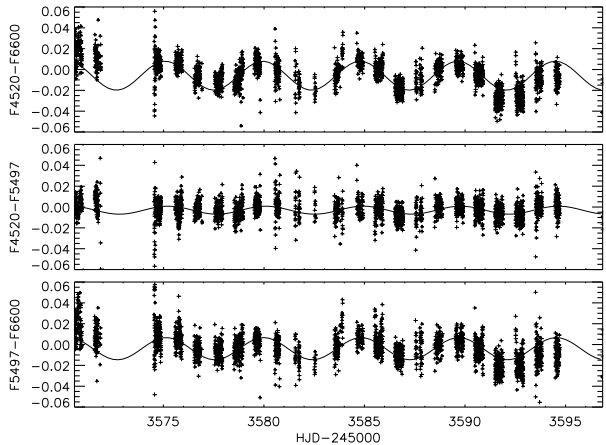,width=3.3truein}
 \caption{Lightcurves of the colors of the starspot variations. F4500-F6600 (top), F4500-F5497 (middle), 
and F5497-F6600 (bottom).  Sine wave fits to the data are overplotted adopting the P$=4.847$~day period
of the modulation.
 }
 \label{fig:spotcolor}
\end{figure}

The amplitude of the variability is a function of passband, indicating the wavelength dependence of 
the flux of the cool starspots causing the variability.  Adopting our derived period, we perform
a linear least squares fit of the phase-folded lightcurve with respect to a sine function to
determine the amplitude of the initial fourier mode.
% and the epoch at zero phase.
The amplitude of the variability is is 0.051~mags, 0.047~mags, and 0.039~mags in the F4520, F5497, and F6600
filters.  The H$\alpha$ band shows an 0.035~mag amplitude variability.
The decrease in amplitude with wavelength is expected if the majority of starspots are
cooler than the stellar photosphere.  
Figure~\ref{fig:spotcolor} shows the colors of the variability emphasizing the change in amplitude of the 
modulation with wavelength.

%The best fit epoch changes systematically with wavelength
%for the continuum bands by as much as 35 minutes.  This could be caused by the structure and position of the starspots.
%In addition, the amplitude of the modulation decreases with increasing wavelength.   
%The starspots are typically cooler than the stellar
%photosphere causing the amplitude of the variability to be largest in the bluest band.  

\subsection{Flare Activity}
\label{sec:flare}

\begin{figure}
 \epsfig{file=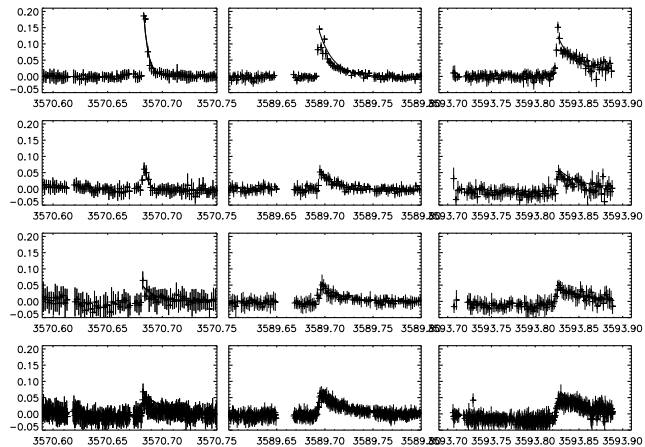,width=3.3truein,angle=0}
 \caption{Regions of the lightcurves including the three flares observed flares.  Flare 1 is shown in the left
column F4500 (top), F5497 (second from top), F6600 (second from bottom), and H$_{\alpha}$ (bottom) filters. 
Flare~2 is in the middle column, and Flare~3 is in the right column.  The best fit exponential models to the
decaying part of each flare are overplotted.
 }
 \label{fig:flare}
\end{figure}

We report the detection of three large optical flares, observed in all four filters, 
during the month long monitoring campaign of AU~Mic.  Three of our filters
show flaring of the continuum of the star; the H$\alpha$ 
filter represents the chromsopheric activity.  The observed
flares show the characteristic structure of a rapid rise time 
and slower decay.  They tend to occur when the star is near
the minimum of the lightcurve which suggests they emanate from the most
heavily spot covered hemisphere of the star.  This has been observed in 
T~Tauri flare stars such as V410 Tau \citep{Fernandez}.
Magnetic fields impede convection on the photosphere 
giving rise to starspots, thus it is not surprising that 
the flares, caused by magnetic activity, are associtated with the heavily covered hemisphere of the star.
%The flares occur at the stellar phases of approximately 0.35, 0.27 and 0.12,
%so they do not emanate from exactly the same active region on the stellar surface.
A close-up of the lightcurve flare regions are shown in Figure~\ref{fig:flare}.

We have measured the peak amplitude of the flares in each filter, as well 
as the rise time and exponential decay timescale. The values are found in Table~\ref{tab:flareprop}.  
The continuum filters show increasing amplitude with decreasing wavelenth.
We resolve the rise time
of the flare in H$\alpha$, but not in any of the other filters due to our
higher H$\alpha$ sampling rate.  However, the data suggest the faster the rise
time of the flares, the faster the decay timescale.  
In addition, the shorter wavelength observations tend to show
faster rise and decay timescales.  Finally, there appears to be pre-flare absorption
in the H$\alpha$ lightcurve which is most prominent before the first flare.  

\begin{table*}
\begin{center}
\caption{Flare Properties}
\label{tab:flareprop}
\begin{tabular}{ccccccc}
\hline\noalign{\smallskip}
Flare  & Filter    & Start & Rise  & Exp Decay &  Peak    & Peak        \\
       &           & Time  & Time  & Timescale &  Exp Fit & Delta Mag   \\
       &           &       & (min) & (min)     &  (mag)   & (mag)     \\
\noalign{\smallskip}
\hline\noalign{\smallskip}
\noalign{\smallskip}
1   &  F4520       & 3570.68091  & $<$2.8    &  6.2  &  0.195  & 0.183  \\
1   &  F5497       & 3570.67993  & 2.5-4.9   &  5.3  &  0.076  & 0.063  \\
1   &  F6600       & 3570.68042  & $<$2.5    &  15.6 &  0.046  & 0.065  \\
1   &  H$\alpha$   & 3570.68140  & 0.7-1.4   &  26.5 &  0.064  & 0.070  \\
2   &  F4520       & 3589.69092  & 2.5-4.9   &  21.3  &  0.138  & 0.147  \\
2   &  F5497       & 3589.69165  & 2.1-4.6   &  20.4  &  0.054  & 0.053  \\
2   &  F6600       & 3589.69043  & 7.0-9.5   &  27.6  &  0.052  & 0.053  \\
2   &  H$\alpha$   & 3589.69189  & 3.9-4.6   &  29.3  &  0.066  & 0.066  \\
3   &  F4520       & 3593.81909  & 7.4-9.8   &  22.1  &  0.101  & 0.150  \\
3   &  F5497       & 3593.82300  & 2.5-5.3   &  30.8  &  0.056  & 0.064  \\
3   &  F6600       & 3593.82007  & 7.7-10.2  &  61.5  &  0.047  & 0.055  \\
3   &  H$\alpha$   & 3593.82275  & 4.2-5.3   &  33.5  &  0.052  & 0.052  \\

\end{tabular}
\end{center}
\end{table*}

\section{Conclusions and Future Work}
\label{sec:concl}

We have obtained unprecedented high cadence, high precision photometry of the young, M1Ve, debris disk star, AU~Mic
in four medium band filters over 28~nights.  The combined lightcurve has a sampling rate
of less than 1~minute and an rms (when the intrinsic stellar variability is removed) of 6.8~mmag.
The lightcurve was searched for transiting extra-solar planets by applying a box-fitting 
algorithm designed to detect gray (same depth in all filters), periodic, square-shaped dips in brightness.
We detected one significant event S/N~$=10.0$ which occured at the end of the night in all 
four filters with a depth$\sim 26.4$~mmag.  However, the constraints placed by the 
depth, period, and shape of the observed dip suggest it is unlikely to be caused by a transiting planet.
More data is required to determine its nature.  Therefore, in the observed lightcurve,
we find no convincing transit events which could be caused by a planet orbiting AU~Mic.

We performed Monte Carlo simulations by adding 
fake transits to the lightcurve to determine the probability of detecting a planet around AU~Mic 
given the window function and noise properties of our data.  The signature of a young Jupiter
mass planet on a short period orbit would be easily detected.  The results of the simulations
indicate there are no planets with masses, M~$> 1$~M$_{J}$ orbiting AU~Mic in the plane of the
debris disk ($\sim 1^{\circ}$) with periods, P~$< 5$~days.  A young Neptune-like planet with 
a smaller radius ($2.5\times$ smaller than the young Jupiter) could also be detected in our data.  
The lack of such a detection indicates there are no young Neptune-like planets orbiting AU~Mic
with periods, P~$< 3$~days.

In addition to the transit search, the high cadence, multi-band photometry is ideal for examining the intrinsic
stellar variability of the star due to starspots and flaring activity.  
AU~Mic exhibits quasi-sinusoidal variability likely due to an uneven distribution of 
starspots on the stellar surface.  The amplitude of the variability varies from 0.051~mags in the blue to 0.039~mags 
in the red.  Using the periodic modulation, we derive a stellar rotation period for the star of 4.847~days.  We also
report on the detection of three large optical flares which tend to emanate from the most heavily spotted
hemisphere of the star.  A more indepth analysis and modelling of the intrinsic stellar variability will
be discussed in a subsequent paper.

New observations that were obtained in August and September of 2006 using the CTIO-1m and Australian
National Universty (ANU) 40-inch telescope will help place further constraints on the existence of
planets around AU~Mic and allow improved modelling of the intrinsic stellar variability.  

\section*{Acknowledgments}

DM and IT are partially supported by Fondap Center for Astrophysics 15010003.  
This research was supported in part by NASA grant NAG5-7697.

\label{lastpage}


\begin{thebibliography}{99}
\bibitem[Aigrain et al.(2007)]{SuzMonitor} Aigrain, S., Hodgkin, 
S., Irwin, J., Hebb, L., Irwin, M., Favata, F., Moraux, E., \& Pont, F.\ 
2007, MNRAS, 375, 29 
\bibitem[Barrado y Navascues et al.(1999)]{Barrado} Barrado et al.\ 1999, ApJL, 520, L123
\bibitem[Baraffe et~al.(1998)]{Baraffe98}
  Baraffe, I., Chabrier, G., Allard, F., \& Hauschildt, P.~H.\ 1998, A\&A, 337, 403
\bibitem[Baraffe et al.(2002)]{Baraffe02} Baraffe, I., Chabrier,
G., Allard, F., \& Hauschildt, P.~H.\ 2002, A\&A, 382, 563
\bibitem[Chen et al.(2005)]{Chen} Chen, C.~H., et al.\ 2005, ApJ, 634, 1372
\bibitem[Clampin et al.(2003)]{Clampin} Clampin, M., et al.\ 2003, AJ, 126, 385
\bibitem[Claret(2000)]{Claret} Claret, A.\ 2000, A\&A, 363, 1081
\bibitem[Fern{\'a}ndez et al.(2004)]{Fernandez} Fern{\'a}ndez,
M., et al.\ 2004, A\&A, 427, 263
\bibitem[Fuhrmeister et al.(2005)]{Fuhr} Fuhrmeister, B.,
Schmitt, J.~H.~M.~M., \& Hauschildt, P.~H.\ 2005, A\&A, 439, 1137
\bibitem[Greaves et al.(2005)]{Greaves} Greaves, J.~S., et al.\ 2005, ApJL, 619, L187
\bibitem[Haisch et al.(2001)]{Haisch01b} Haisch, K.~E., Jr.,
Lada, E.~A., \& Lada, C.~J.\ 2001, ApJL, 553, L153
\bibitem[Holland et al.(2003)]{Holland} Holland, W.~S., et al.\ 2003, ApJ, 582, 1141
\bibitem[Irwin \& Lewis(2001)]{IrwinLewis} Irwin, M. \& Lewis, J. 2001, NewAR, 45, 105
\bibitem[Kalas et al.(2004)]{Kalasa} Kalas et al.\ 2004, Science, 303, 1990
\bibitem[Kov\'acs et~al.(2002)]{Kovacs} Kov\'acs, G., Zucker, S., \& Mazeh, T. 2002, A\&A, 391, 369
\bibitem[Krist et al.(2005)]{Krist} Krist, J.~E., et al.\ 2005, AJ , 129, 1008 
\bibitem[Liu et al.(2004)]{Liua} Liu et al.\ 2004, ApJ , 608, 526
\bibitem[Liu(2004)]{Liub} Liu, M.\ 2004, Science 305, 1442 
\bibitem[Mandel \& Agol(2002)]{MandelAgol} Mandel, K., \& Agol, E.\ 2002, ApJL, 580, L171
\bibitem[Masciadri et al.(2005)]{Masc} Masciadri, E., Mundt,
R., Henning, T., Alvarez, C., \& Barrado y Navascu{\'e}s, D.\ 2005, ApJ, 625, 1004
\bibitem[Metchev et al.(2005)]{Metchev} Metchev, S.~A., Eisner,
J.~A., Hillenbrand, L.~A., \& Wolf, S.\ 2005, ApJ, 622, 451
\bibitem[Paulson et al.(2006)]{Paulson} Paulson, D.~B., Allred,
J.~C., Anderson, R.~B., Hawley, S.~L., Cochran, W.~D., \& Yelda, S.\ 2006,
PASP, 118, 227
\bibitem[Perryman et al.(1997)]{Perryman} Perryman, M.~A.~C., et al., 1997,
The Hipparcos and Tycho Catalogues, ESA SP-1200.   
\bibitem[Pont et al.(2006)]{Pont06} Pont, F., Zucker, S., \&
Queloz, D.\ 2006, MNRAS, 1146
\bibitem[Quillen \& Thorndike(2002)]{QuillThorn} Quillen, A.~C.,
\& Thorndike, S.\ 2002, ApJL, 578, L149
\bibitem[Roberge et al.(2005)]{Roberge} Roberge, A.,
Weinberger, A.~J., Redfield, S., \& Feldman, P.~D.\ 2005, ApJL, 626, L105
\bibitem[Rodono et al.(1986)]{Rodono} Rodono, M., et al.\
1986, A\&A, 165, 135
\bibitem[Scargle(1982)]{Scarg} Scargle, J.~D.\ 1982, ApJ,
263, 835
\bibitem[Strom \& Edwards(1993)]{Strom93} Strom, S.~E., \&
Edwards, S.\ 1993, ASP Conf.~Ser.~ 36: Planets Around Pulsars, 36, 235
\bibitem[Tingley \& Sackett(2005)]{TingSack} Tingley, B., \& 
Sackett, P.~D.\ 2005, ApJ, 627, 1011 
\bibitem[Torres et al.(1972)]{Torres} Torres, C.~A.~O., Mello,
S.~F., \& Quast, G.~R.\ 1972, ApJL, 11, 13
\bibitem[Weinberger et al.(2003)]{Weinberger} Weinberger, A.~J.,
Becklin, E.~E., \& Zuckerman, B.\ 2003, ApJL, 584, L33
\bibitem[Wyatt \& Dent(2002)]{WyattDent} Wyatt, M.~C., \& Dent,
W.~R.~F.\ 2002, MNRAS, 334, 589
\bibitem[Wyatt et al.(1999)]{Wyatt99} Wyatt, M.~C., Dermott,
S.~F., Telesco, C.~M., Fisher, R.~S., Grogan, K., Holmes, E.~K., \&
Pi{\~n}a, R.~K.\ 1999, ApJ, 527, 918
\end{thebibliography}
\end{document}